*Type of the Paper : Original Research*

# A Holistic Approach on Smart Garment for Patients with Juvenile Idiopathic Arthritis

**Safal Choudhary[1], Princy Randhawa[2], Sampath Kumar J.P [3] *, Shiva Prasad H.C [4],***

1. Department of Fashion Design, Manipal University Jaipur, Jaipur , Rajasthan 303007; India safal220651001@muj.manipal.edu
2. Department of Mechatronics Engineering, Manipal University Jaipur, Jaipur, Rajasthan 303007; India princy.randhawa@jaipur.manipal.edu
3. Department of Fashion Design, SoDA, FoD, Manipal University Jaipur, Jaipur , Rajasthan 303007; India sampath.kumar@jaipur.manipal.edu
4. Department of Mechanical Engineering, SAMM, Faculty of engineering, Manipal University Jaipur, Jaipur, Rajasthan 303007; India shiva.prasad@jaipur.manipal.edu

**\*Correspondence:** shiva.prasad@jaipur.manipal.edu; sampath.kumar@jaipur.manipal.edu

**Abstract:**

Globally, arthritis is a prevalent chronic condition affecting numerous individual. The juvenile idiopathic arthritis (JIA) is characterized by joint inflammation leading to pain, swelling, stiffness, and limiting the movements and is the dominant type of arthritis found in children and adolescents. Individuals suffering from JIA requires an ongoing treatment for their lifetime. Beyond inflammation's, the JIA patients expressed their concerns about various factors and the lack of responsive services addressing their challenges. One of featured design is wearble garments that helps them to do daily activites. To address this issues, the research aims to design a smart garment with holistic approach. By providing smart wearble garment, and helps to improve well-being, coping abilities, independence, and overall quality of life for patients. The results shows with juvenile idiopathic arthritis.

## 1. Introduction

In the medical landscape, Juvenile Idiopathic Arthritis (JIA) stands as the prevalent chronic rheumatic disease in children and a significant contributor to both short-term and long-term disability (Ludovico Di et al. 2023) [1]. JIA is a chronic rheumatic disease of unknown origin, with a higher occurrence among young female patients (Barut K, Adrovic A, Şahin S, Kasapçopur Ö, 2017)[2]. The International League Against Rheumatism (ILAR) identifies seven distinct and mutually exclusive categories of JIA, determined by the disease manifestations observed within the first 6 months of its onset (Martini, A., Lovell, D.J., Albani, S. *et al.* 2022)[3], the subtypes are depicted in Fig 1.  JIA encompasses all forms of chronic childhood arthritis, affecting not only joints but also extra-articular structures, which can lead to disability and, in severe cases, even associated fatality (Zaripova, L.N., Midgley, A., Christmas, S.E. et al. 2021) [4]. As active JIA persists into adulthood, the cumulative effect leads to a higher degree of functional limitation and joint destruction. Nevertheless, some patients may encounter detrimental effects such as joint deformities, destruction, growth abnormalities, and retardation, leading to pain, impaired psychological health, or challenges with daily activities (J. C. Packham , M. A. Hall, 2002) [5].

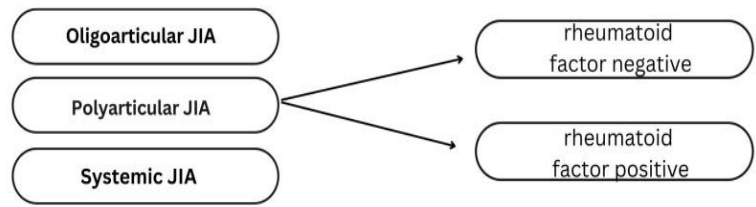

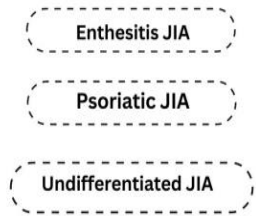

Fig 1 *Types of Juvenile idiopathic arthritis*

The diagnosis is challenging as there are no specific tests to definitively confirm the condition (Capurso [6] et al., 2016). Around 5% of children with JIA exhibit rheumatoid factor (RF)–positive arthritis, which closely resembles adult rheumatoid arthritis (RA) in its phenotype. Notably, the presence of rheumatoid factor persists throughout the patient's lifetime (Anne Hinks, 2018) [7]. For patients who travel and live away from home for educational or work purposes, it can be particularly challenging. Activities that impact arthritic patients include running, cycling, dancing, traveling, climbing, dressing, skipping, engaging in physical games, lifting bags, writing, typing, opening jars and door knobs, as well as chopping food. Additionally, Arthritis sufferers may experience a detrimental effect on their sleep quality, and it is equally important to acknowledge that inadequate sleep can worsen pain levels and heighten stress levels (Grabovac I et el. 2018)[8]. The impact extends beyond inflammation, disrupting their daily routines and requiring more responsive services (Chaplin et al. 2018)[9]. In light of this, the importance of smart garments to their needs cannot be overstated. A smart garment can be described as an intelligent system with the capability to sense and interact with the wearer's surroundings, conditions, and stimuli (Tao, X.,2001 [10], Gilsoo et al. 2009 [11]). Smart garments possesses the ability to remotely predict, prevent, and monitor chronic illnesses, thereby decreasing the necessity for hospital stays and empowering patients to maintain their independence (Brown, D. 2020 [12]). Smart garments are specifically crafted to cater to health and containing various garments such as T-shirts (Sankaran [13], Stojanović [14], Perego [15] , bras (Moreno 2019 [16], Elsheakh 2023 [17], Raji 2020 [18], Navalta 2020 [19], Lee 2021 [20] ), sleeves (Feng 2023 [21] Qiao 2022 [22], DelPreto 2023 [23] , jackets (Vuohijoki 2022 [24], Azam Khan 2023 [25], Uzun [26]), leggings (Hwang [27], Bravo [28], Barhoumi [29]), and more (Rudolf [30], with a focus on healthcare applications. Considering this, the importance of smart garments for their needs cannot be emphasized enough. These essential resources play a crucial role in empowering patients with JIA to navigate their condition more effectively and maintain their independence, ensuring a better quality of life. This article focuses on exploring the requirements of patients having Juvenile Idiopathic Arthritis (JIA), specifically in the context of designing smart garment which a patient can use as a sleepwear also. The research aims to delve into the unique needs of these individuals and understand how to design smart garment to accommodate their condition. Research aims to raise awareness and promote further research and development in this crucial area of care, to ensure that patients with JIA have access to the resources they need to lead fulfilling lives and overcome the obstacles posed by their condition.

## 2. Objective

The objective of this study is to identify and understand the specific needs of Juvenile idiopathic arthritic patients regarding smart garment. The focus is on exploring their requirements, preferences, and challenges related to different garments. By gathering insights directly from this demographic group, the study aims to inform the development of tailored solutions that address their unique needs and empower them to engage in daily activities with greater ease and confidence.

## 3. Literature Review

Research done in order to develop design requirements for functional sports bras for young women with juvenile idiopathic arthritis, to help retailers and academia better understand their needs (Mccoy L. 2021) [31]. Adaptive clothing empowers woman with rheumatoid arthritis, they can dress up independently and comfortably and still feel good about herself, which increases self-confidence. In this study, the researcher devised the patterns of kurta using pattern making techniques. (Shalini Singh 2019) [32]. The restricted joint functions of rheumatoid arthritis (RA) patients pose challenges in dressing, particularly with certain garments like sarees. In this study, alternative options are proposed for easier-to-drape sarees tailored to arthritis patients' needs (Chandra [33], et al. 2016). Adaptive clothing research found special features to be highly practical, Three sets of adaptive garments were designed for females with rheumatoid impairements, independence and self-confidence were assessed by wear trail test (Naheed Azher, et al. 2012)[34]. The study's main goal is to empower elderly long-term care patients with dressing independence by devising alternative fastener options for dresses. (L. Sperling, M. Karlsson 1989)[35]. This research addresses Rheumatoid arthritis, the proposed workflow concentrates on design requirements such as breathability, support in affected areas, lightweight design, and customization to alleviate individual pain points through a wearable glove solution (Jierui Fang 2020)[36]. A review assessed therapy gloves' effectiveness in hand function for RA patients, in this review grip strength, pinch strength, ROM and dexterity were selected as the outcome measures to check the effectiveness of therapy gloves (Nasir SH, Troynikov O, Massy-Westropp N, 2014)[37]. The study introduces a Bluetooth-enabled glove that quantitatively assesses hand performance for rheumatoid arthritis patients, providing an alternative to subjective descriptions. Additionally, the glove is designed with rubber adjustable rings for a comfortable fit (Socher R. 2018)[38]. The study explores wrist orthotics typically crafted from Neoprene material, known for its resistance to abrasion, waterproof properties, and stretchability. These orthotics include five separate hook and close straps for secure fastening and a palmar side pocket to hold a metal bar, providing crucial wrist support (Underwood, S. 2009)[39].

Table 1. Studies related to arthritis garments and accessories

| References | Garment Type | Nature of work | Type of patient | Limitations |
|---|---|---|---|---|
| Mccoy (2021). | Sports Bra | Developing sports bras for young arthritic women: Required features and functions. Easy joint movement (Adaptive garment) | Young adolescent girls, Juvenile Idiopathic Arthritis | The study does not incorporate smart technology to assist women in wearing sports bras for shoulder, neck, and back pain. Prototype and trial test was missing in research. |
| Singh (2019). | Kurta pattern design | A Study on Adaptive Clothing for Females with Arthritis (Adaptivegarment) | Old women, Rheumatoid arthritis | Limited geographical representation of participants. The research lacked. |
| Chandra, & Anand, (2016). | Saree design | Adaptive clothing for the Rheumatoid arthritis patients, Easy to drape saree | Women, Type of JIA Rheumatoid arthritis | Limited geographical representation of participants, trial test was missing in research. |
| Azher (2012). | Three set clothing | Adaptive clothing for rheumatoid impairemnet | Women with rheumatoid impairement | Only focused on the enthnic wear. Limited demographic approach used in the study. |
| Sperling & Karlsson (1989). | Clothing fastners | Clothing fasteners for long-term-care patients: Evaluation of standard closures and prototypes on test garments | Both Elderly men and women, Osteoarthritis, Rheumatoid | Small sample size, limited diversity. |
| Fang (2020). | Glove | Responsive wearables for rheumatoid arthritis. Customization for individual pain point, support in affected area (Smart wearable) | Both men and women, Type of JIA Rheumatoid arthritis | Wearing gloves in varying seasons is not feasible. |
| Socher (2018). | Smart Glove | Wearable Smart Glove Offering Custom Support for People with Rheumatoid Arthritis, monitor the hand movement and provide sport | Both men and women, Rheumatoid arthritis | Wearing gloves across various seasons is not viable. |
| Underwood (2009). | Glove | Smart clothing and disability: wearable technology for people with arthritis. | Both men and women, Rheumatoid arthritis | The glove lacks the capability to track hand activities and provide feedback to the patient and doctor. |

**4. Problem Statement**

The problem addressed in this study revolves around the lack of smart garment for juvenile idiopathic arthritic patients. Despite the existence of general adaptive clothing options, there is a lack of holistic solutions that cater to the unique needs of this demographic group.

**5. Methodology**

The methodology used is ……………………………………………….

**6. Observation and Interviews**

After conducting interviews and observing juvenile idiopathic arthritis (JIA) patients, it became evident that their limited range of motion affects their elbow, wrist, shoulder, knee and hip areas. Consequently, dressing and other essential activities like travelling, using stairs, bicycle riding, become challenging, causing additional suffering due to joint inflammation and stiffness. However, upon introducing the concept of smart adaptive garments to the arthritic patients and their parents, it was observed that they expressed a keen interest in adopting these new ideas to improve their health and functionality. Data collected by using semi structured interview and …………………………………………..

**7. Feasiblity Analysis**

To create a comprehensive framework for this research, the Darden Design Thinking process was utilized in the research, incorporating a user-centred design approach within a sequential mixed methods methodology. The mixed-methods sequential explanatory design consists of two distinct phases: quantitative followed by qualitative (Creswell et al. 2003)[40]. The study consisted of three stages, involving focus group surveys, interviews, and market research on arthritic patients garments. Data collection methods included quantitative and qualitative surveys among patients with JIA. Factor and Effect Analysis was used to identify user requirements, while a postures examination provided insights into participants' physical mobility. Based on the findings, specific design elements and strategies for arthritis-targeted garments were proposed using Clo3D. The research takes a holistic approach with the aim of developing smart garment specifically designed for individuals with Juvenile Idiopathic Arthritis (JIA). This approach involves considering the various aspects and needs of individuals with JIA, beyond just the physical challenges they face. The approach considers the diverse needs of individuals with JIA, addressing physical limitations while prioritizing overall well-being, comfort, and fashion preferences. By acknowledging challenges beyond physical symptoms, including self-esteem, independence, health and social inclusion, the smart adaptive street-wear aims to empower individuals with JIA. The garments feature easy dressing, gentle fabrics, adjustable closures, curative qualities and aesthetically appealing designs to enhance daily life and promote a sense of normalcy and inclusivity (Fig. 2). The holistic approach seeks to improve their quality of life, allowing confident participation in daily activities and social interactions.

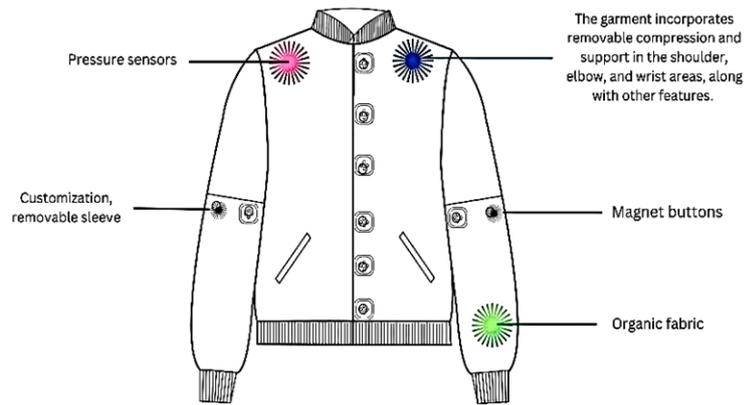

Fig. 2 *Proposed Smart street-wearable design of the upper garment for arthritic young women*

8. Conclusions

The research on requirement for smart garments for arthritic patients has shed light on the significant challenges faced by affected individuals. JIA's impact on joint inflammation, stiffness and limited range of motion affects multiple joints, leading to difficulties in dressing and other essential activities causing additional suffering. However, this research contributes to a better understanding of the specific needs of JIA patients and paves the way for improved interventions and services catering to their requirements. As we move forward, the adoption and implementation of smart garments offer a brighter outlook for the lives of these individuals, fostering a future of improved health, functionality, and enhanced well-being.

Limitations: While the research has provided valuable insights into the challenges faced by individuals with JIA, there are certain limitations to be acknowledged. The study's sample size, geographical diversity, and duration may influence the generalizability of the findings.

Future Scope: As medical and technological advancements continue, there is potential for further research and innovation in the development of adaptive clothing and wearable solutions for individuals with JIA. The integration of advanced materials, sensors, and data analytics can lead to more sophisticated garments that provide real-time monitoring and personalized support, optimizing the management of JIA symptoms and enhancing daily activities for the patients.

**Author Contributions:**

| Safal Choudhary | Conceptualizing, Literature review, and paper draft |
| --- | --- |
| Dr. Princy Randhawa | Collection of Data and formatting |
| Dr. Sampath Kumar J.P. | Statistical calculation and analysis |
| Dr. Shiva Prasad HC | Corresponding author and type setting |

**Funding:** NO
**Institutional Review Board Statement:** APPROVED
**Informed Consent Statement:** YES

**Data Availability Statement:**
**Acknowledgments:**
**Conflicts of Interest:**

**References**

[1] Di Ludovico, A., La Bella, S., Di Donato, G. et al. The benefits of physical therapy in juvenile idiopathic arthritis, 2023, Rheumatol Int, Volume [43], (pp. 1563–1572).

[2] Barut K, Adrovic A, Şahin S, Kasapçopur Ö. Juvenile Idiopathic Arthritis. *Balkan Med J.,* 2017, *Volume* 34[2], (pp. 90-101).

[3] Martini, A., Lovell, D.J., Albani, S. *et al.* Juvenile idiopathic arthritis. *Nat Rev Dis Primers,* 2022, Volume [**8**], (p. 5).

[4] Zaripova LN, Midgley A, Christmas SE, Beresford MW, Baildam EM, Oldershaw RA. Juvenile idiopathic arthritis: from aetiopathogenesis to therapeutic approaches. *Pediatr Rheumatol*, 2021, Volume 19, (p.135).

[5] J. C. Packham , M. A. Hall, Long-term follow-up of 246 adults with juvenile idiopathic arthritis: functional outcome, *Rheumatology*, 2002, *Volume* [41], Issue 12, (pp.1428–1435).

[6] Michele Capurso, Maria Lo Bianco, Elisabetta Cortis & Corrado Rossetti (2016) Constructing an Explanation of Illness with Children: A Sample Case Study of Juvenile Arthritis, Child *Care in Practice*,*Volume* [22:3], (pp. 247-256).

[7] Anne Hinks et al. Brief Report: The Genetic Profile of Rheumatoid Factor–Positive Polyarticular Juvenile Idiopathic Arthritis Resembles That of Adult Rheumatoid Arthritis, *Arthritis and Rheumatology*, 2018, *Volume* [70], Issue 6, (pp. 957-962).

[8] Grabovac I, Haider S, Berner C, Lamprecht T, Fenzl KH, Erlacher L, Quittan M, Dorner TE. Sleep Quality in Patients with Rheumatoid Arthritis and Associations with Pain, Disability, Disease Duration, and Activity. J Clin Med. 2018, *Volume* 7[10], ( p. 336).

[9] Chaplin, Hema & Ioannou, Y. & Sen, D. & Lempp, Heidi & Norton, Sam., OP0101 Exploring pain and the impact of jia on adolescents and young adults: a mixed-methods study. *Annals of the Rheumatic Diseases.*, 2018, [77], (pp. 101.1-101).

[10] Tao, X., Smart technology for textiles and clothing. Edited by Xiaoming Tao. Woodhead Publishing, 2001.

[11] Gilsoo, C., Seungsin, L. and Jayoung, C., *Review and Reappraisal of Smart Clothing*, International Journal of Human-Computer Interaction, Volume [25], (pp. 582-617), 2009. *(PDF) Review of Smart Clothing with Emphasis on Education and Training*

[12] Brown, D. The Arrival Of 2020 Brings Smart Clothing To The Forefront Of Healthcare, YouAreUNLTD.com

[13] Sankaran, Sakthivel & Britto, Preethika & Petchimuthu, Priya & Sushmitha, M. & Rathinakumar, Sagarika & Mallaiyan, Vijay & Ayyavu, Selva. (2023). Monitoring of Physiological and Atmospheric Parameters of People Working in Mining Sites Using a Smart Shirt: A Review of Latest Technologies and Limitations. 10.1007/978-981-19-8493-8_53.

[14] Stojanović, Sandra & Gersak, Jelka & Uran, Suzana. (2022). Development of the Smart T-Shirt for Monitoring Thermal Status of Athletes. Autex Research Journal. 23. 10.2478/aut-2022-0005.

[15] Perego, Paolo & Sironi, Roberto & Gruppioni, Emanuele & Andreoni, Giuseppe. (2023). TWINMED T-SHIRT, a Smart Wearable System for ECG and EMG Monitoring for Rehabilitation with Exoskeletons. 10.1007/978-3-031-35741-1_40.

[16] Moreno, Marie-Valerie & Herrera, Edouard. (2019). Evaluation on Phantoms of the Feasibility of a Smart Bra to Detect Breast Cancer in Young Adults. Sensors (Basel, Switzerland). 19. 10.3390/s19245491.

[17] Elsheakh, Dalia & Elgendy, Yasmine & Elsayed, Mennatullah & El-Damak, A.R.. (2023). Circularly Polarized Textile Sensors for Microwave-Based Smart Bra Monitoring System. Micromachines. 14. 586. 10.3390/mi14030586.

[18] Raji, King & Miao, Xuhong & Wan, Ailan & Niu, Li & Li, Yutian & Boakye, Andrews. (2020). Design and Fabrication of Smart Bandeau Bra for Breathing Pattern Measurement. 10.1007/978-3-030-32523-7_3.

[19] Navalta, James & Ramirez, Gabriela & Maxwell, Crystal & Radzak, Kara & McGinnis, Graham. (2020). Validity and Reliability of Three Commercially Available Smart Sports Bras during Treadmill Walking and Running. Scientific Reports. 10. 10.1038/s41598-020-64185-z.

[20] Lee, Suhyun & Rho, Soo & Lee, Sojung & Lee, Jiwoong & Lee, Jiho & Lim, Daeyoung & Jeong, Wonyoung. (2021). Implementation of an Automated Manufacturing Process for Smart Clothing: The Case Study of a Smart Sports Bra. Processes. 9. 289. 10.3390/pr9020289.

[21] Feng, Yan & Wang, Hao-xiang & Liu, Peng-bin & Qi, Hua & Pan, Rui-zhi & Zhang, Hong-pu & Zhang, Hua. (2023). Optical Fiber Bragg Grating Based Sensing System of Flexible Wearable Smart Sleeve for Tracking Human Arm Joint Movements. Measurement Science and Technology. 34. 10.1088/1361-6501/acd4d7.

[22] Qiao, Guofu & Sun, Jiongfeng. (2022). Self-monitoring of stresses in grouted sleeves using smart grout. Structural Health Monitoring. 21. 147592172210805. 10.1177/14759217221080516.

[23] DelPreto, Joseph & Brunelle, Cheryl & Taghian, Alphonse & Rus, Daniela. (2023). Sensorizing a Compression Sleeve for Continuous Pressure Monitoring and Lymphedema Treatment Using Pneumatic or Resistive Sensors. 10.1109/RoboSoft55895.2023.10122068.

[24] Vuohijoki, Tiina & Ihalainen, Tiina & Merilampi, Sari & Virkki, Johanna. (2022). Multidisciplinary development of Smart Jacket for elder care. Finnish Journal of eHealth and eWelfare. 14. 10.23996/fjhw.112214.

[25] Azam Khan, Aman Ul & Kumar Saha, Aurgho & Bristy, Zarin & Tazrin, Tasnima & Baqui, Abdul & Dev, Barshan. (2023). Development of the Smart Jacket Featured with Medical, Sports, and Defense Attributes using Conductive Thread and Thermoelectric Fabric. 10.3390/engproc2023030018.

[26] Uzun Ozsahin, Dilber & Almoqayad, Abdulrahim & Ghader, Abdullah & Alkahlout, Hesham & Idoko, John & Duwa, Basil & Ozsahin, Ilker. (2022). Development of smart jacket for disc. 10.1016/B978-0-323-85413-9.00003-7.

[27] Hwang, Jin-Hee & Jee, Seunghyun & Kim, Sun. (2022). Developing a Prototype of Motion-sensing Smart Leggings. Fashion & Textile Research Journal. 24. 694-706. 10.5805/SFTI.2022.24.6.694.

[28] Bravo Carrasco, Valeria Paz & Vidal, Javier & Caparrós-Manosalva, Cristián. (2022). Vibration motor stimulation device in smart leggings that promotes motor performance in older people. Medical & Biological Engineering & Computing. 61. 10.1007/s11517-022-02733-7.

[29] Barhoumi, Houda. (2021). A novel design approach and ergonomic evaluation of Class I compression legging. International Journal of Clothing Science and Technology. ahead-of-print. 10.1108/IJCST-11-2020-0179.

[30] Rudolf, Andreja & Stjepanovič, Zoran & Penko, Tadeja. (2022). Review of Smart Clothing with Emphasis on Education and Training. 10.2478/9788366675735-036.

[31] Mccoy, L., Developing sports bras for young arthritic women: required features and functions, 2021.

[32] Singh, Shalini., A Study on Adaptive Clothing for Females with Arthritis. *International Journal of Advanced Scientific Research and Management,* 2019, Volume [5], (p. 4).

[33] Chandra, Ruchi & Anand, Noopur. Adaptive clothing for the rheumatoid arthritis patients, *International textiles & apparel sustainability conference held in Mauritius Adaptive Clothing for The Rheumatoid Arthritis Patients*, 2016.

[34] Naheed Azher, & Muhammad Saeed. (1). Adaptive Clothing for females with arthritis impairement. *Journal of University Medical & Dental College*, 2012, *3*[2], (pp. 52-59).

[35] L. Sperling, M. Karlsson,Clothing fasteners for long-term-care patients: Evaluation of standard closures and prototypes on test garments,Applied Ergonomics, 1989,*Volume* [20], Issue 2,1989, (pp. 97-104).

[36] Jierui Fang, Responsive Wearables for Rheumatoid Arthritis, *Massachusetts Institute of Technology*, 2020.

[37] Nasir SH, Troynikov O, Massy-Westropp N. Therapy gloves for patients with rheumatoid arthritis: a review. Ther Adv Musculoskelet Dis., 2014, Volume 6[6], (pp. 226-237).

[38] Ramona Socher, "MANOVIVO: Wearable Smart Glove Offering Custom Support for People with Rheumatoid Arthritis," *Wearable Technologies*, 2018.

[39] Underwood, S., Smart clothing and disability: wearable technology for people with arthritis. *Smart Clothes and Wearable Technology,* 2009, *Volume [4],* (pp. 371–387).

[40] Creswell, J. W.,Research design: Qualitative, quantitative, and mixed methods approaches (2nd ed.). 2003, *Thousand Oaks*, CA: Sage.